\documentclass[IOP,preprint,ppcf]{revtex4-1}
\usepackage{amssymb}
\usepackage{amsmath}
\usepackage{graphicx}
\usepackage{bm}
\usepackage{ulem}
\usepackage{color}
\usepackage{mathrsfs}
\usepackage{subfigure}
\usepackage{caption}
\usepackage{epstopdf}
\usepackage{CJK}
\DeclareMathOperator{\sgn}{sgn}

\DeclareMathOperator{\Imag}{Im}

\draft

\begin{document}


\title{Analytical model for quasi-linear flow response to resonant magnetic perturbation in resistive-inertial and viscous-resistive regimes}

\author{Wenlong Huang}
\affiliation{School of Computer Science and Technology,\\
Anhui  Engineering Laboratory for  Industrial Internet  Intelligent Applications and Security, Anhui University of Technology, Ma'anshan, Anhui 243002, China}

\author{Ping Zhu}
\email[E-mail:]{zhup@hust.edu.cn}
\affiliation{International Joint Research Laboratory of Magnetic Confinement Fusion and Plasma Physics, State Key Laboratory of Advanced Electromagnetic Engineering and Technology, School of Electrical and Electronic Engineering, Huazhong University of Science and Technology, Wuhan, Hubei 430074, China\\
Department of Engineering Physics, University of Wisconsin-Madison,
Madison, Wisconsin 53706, USA
}

\author{Hui Chen}
\affiliation{Department of Physics, Nanchang University, Nanchang, Jiangxi 330031, China}

\date{\today}

\begin{abstract}
In this work, a quasi-linear model for plasma flow response to the resonant magnetic perturbation (RMP) in a tokamak has been rigorously developed in the resistive-inertial (RI) and viscous-resistive (VR) regimes purely from the two-field reduced MHD model. Models for plasma response to RMP are commonly composed of equations for the resonant magnetic field response (i.e. the magnetic island) and the torque balance of plasma flow. However, in previous plasma response models, the magnetic island and the torque balance equations are often derived separately from reduced MHD and full MHD equations, respectively. By contrast, in this work we derive both the magnetic island response and the torque balance equations in a quasi-linear model for plasma flow response entirely from a set of two-field reduced MHD equations. Such a quasi-linear model can recover previous plasma flow response models within certain limits and approximations. Furthermore, the physical origins of quasi-linear forces and moments in the flow response equation are also accurately calculated and clarified self-consistently.

\end{abstract}

\maketitle

\section{Introduction}
Resonant Magnetic Perturbation (RMP) coils have been widely equipped in fusion devices due to their emerging and promising potential for controlling plasma properties and behaviors~\cite{jak09, paz15, wang16a, ding2018, liang2019, yang19a}. For example, experiment and simulation results in J-TEXT~\cite{zhuang11} show that RMP coils can be employed to control tearing modes and runaway electron activities~\cite{hu13a, chen18a}. In the last decades, edge localized modes (ELMs) suppression and mitigation by RMPs have been realized in various tokamaks~\cite{evans05a, sut11a, sun16}.


It is believed that the mechanism of ELM suppression or mitigation by RMP coils is closely connected to the plasma response to external magnetic perturbations~\cite{liu10, fitz14, fitz18, fitz2019, hu19}. Previous theory models on error field are often directly and heuristically applied to plasma response in both the viscous-resistive and the Rutherford regimes~\cite{fitz93a,fitz14}, which have recently been extended to include the two-fluid and neo-classical flow effects~\cite{fitz18, fitz2019}. Predictions from those theory models are highly relevant to the interpretation of experimental results on the RMP-induced ELM suppression (e.g.~\cite{fitz93a, fitz14, hu19}). Nonetheless, most previous theory models are constructed mainly on heuristic bases instead of more rigorous or self-consistent derivations (e.g.~\cite{fitz14,huang19}). For example, in those models, the island evolution equation for nonlinear plasma response is derived from reduced MHD model, whereas the torque balance equation is a direct outcome of the full MHD equations. 


In this work, we propose a more self-consistent approach to the derivation of the plasma flow response model in both the resistive-inertial (RI) and viscous-resistive (VR) regimes in cylindrical geometry within the framework of the two-field reduced MHD equations. The model is composed of the plasma response equation and the poloidal angular momentum equation in the spectral space of Bessel functions. By absorbing the rigid time-dependent flow into boundary perturbation and dropping the quasi-linear magnetic terms, we extend our previous linear plasma response solutions in slab configuration~\cite{huang19} to cylindrical geometry in presence of rigid time-dependent poloidal flow for both RI and VR regimes, which can reduce to the earlier solutions in the case of steady state flow, as well as the earlier steady state solutions of linear plasma responses in corresponding regimes for the same assumptions~\cite{fitz91a}. The extension of linear plasma response solutions to allow the presence of time-dependent in addition to steady state flow, enables the construction of quasi-linear stresses that are more self-consistent with the dynamic nature of plasma flow in the plasma momentum equation. After obtaining the linear plasma response solutions, we further expand the poloidal angular momentum equation including the quasi-linear stresses in the Bessel spectral space, where the quasi-linear forces retain the Maxwell torque without any assumption on its radial profile~\cite{fitz93a}. Under certain approximations, the newly derived torque balance equation can naturally reduce to its less complete versions in Refs.~\cite{huang19, fitz2019}. The new derivation also allows us to clarify the physical meanings of the quasi-linear forces and moments~\cite{cole15a}.

It should be noted that the toroidal flow cannot be rigorously considered within the two-field reduced MHD framework adopted in this work. To study the dynamics of toroidal flow, at least the four-field reduced MHD mode or full MHD model should be used. Also, the quasi-linear magnetic effects are neglected here, which should be included in the highly nonlinear regime where the magnetic island width is much larger than the resistive tearing layer width. Nonetheless, the self-consistent quasi-linear plasma flow response model within the framework of two-field reduced MHD equations should provide a solid foundation to the building of the plasma response model including toroidal flow and quasi-linear magnetic effects next.


The rest of the paper is organized as follows. In Sec. II, we introduce the reduced MHD model in the cylindrical geometry. In Sec. III, the linear plasma response solutions in the RI and VR regimes with time-dependent flow are obtained. Meanwhile, we derive the poloidal angular momentum equation in the Bessel spectral space and construct the relevant plasma flow response model in Sec. IV. Finally, a summary and discussion is given in Sec. V.

\section{Two-field reduced MHD model}

%

In the cylindrical coordinate system ($r$, $\theta$, $z$), we consider the plasma response to external magnetic field perturbation in a low-$\beta$, large aspect ratio, periodic ``straight'' tokamak equilibrium. Introducing the flux function $\psi$ and stream function $\phi$, the magnetic field and velocity can be written as $\vec{B} = B_z\vec{e}_z + \vec{e}_z\times\nabla\psi$, where $B_z$ is the constant toroidal magnetic field, and $\vec{v} = \vec{e}_z\times\nabla\phi$. In the low $\beta$ plasma, the perturbed toroidal components of magnetic field and velocity can be neglected. Then, the incompressible two-field reduced MHD model governing $\psi$ and $F$ are given, respectively, by\cite{xu15}
\begin{eqnarray}
&\frac{\partial\psi}{\partial t}+(\vec{e}_z\times\nabla\phi)\cdot\nabla\psi - B_z\partial_z\phi = \eta j_z \label{rmhd1},\\
&\rho(\frac{\partial}{\partial t}+\vec{v}\cdot\nabla)F=\vec{B}\cdot\nabla j_z + \nu_\perp \nabla^2 F \label{rmhd2},
\end{eqnarray}
where $\rho$, $\eta$, and $\nu_\perp$ are plasma density, resistivity, and viscosity, respectively. In addition, the vorticity
\[
F = \vec{e}_z\cdot\nabla\times\vec{v} = \nabla_\perp^2\phi = \left[\frac{1}{r}\frac{\partial}{\partial r}(r\partial_r) + \frac{\partial_\theta^2}{r^2}\right]\phi,
\]
and the toroidal component of current density
\[
j_z=\vec{e}_z\cdot\vec{j} = \frac{1}{\mu_0}\nabla_\perp^2\psi = \frac{1}{\mu_0}\left[\frac{1}{r}\frac{\partial}{\partial r}(r\partial_r) + \frac{\partial_\theta^2}{r^2}\right]\psi.
\]

In the cylindrical geometry, any quantity can be written as $f = f_{\rm eq} + \delta f$, where $f_{\rm eq} = f_{\rm eq}(r)$ and $\delta f = \delta f(r, \theta, \phi, t)$ are the equilibrium and perturbation parts of $f$, respectively. We expand the perturbed quantity as $\delta f= \sum_{l=-\infty}^{\infty}\delta f_l e^{il(m\theta- n\frac{z}{R_0})}$, where $m$ ($n$) is the poloidal (toroidal) mode number, and
$a$ ($R_0$) is the minor (major) radius of plasma. Adopting the single helicity approximation, neglecting the quasi-linear magnetic terms, and keeping only the quasi-linear flow effects, Eqs.~\eqref{rmhd1} and \eqref{rmhd2} can be reduced to the following
\begin{align}
&\partial_t \delta\psi_1 + \delta\bm{v}_1\cdot\nabla\psi_{\rm eq} - B_z\partial_z\delta\phi_1 +  \bm{v}_0\cdot\nabla\delta\psi_1 = \eta \delta j_{z1},\label{psi1linear}\\
&\rho[\partial_t \delta F_1 + \bm{v_{0}}\cdot\nabla \delta F_1 + \delta\bm{v_{1}}\cdot\nabla F_0] = \bm{B}_{\rm eq}\cdot\nabla \delta j_{z1} + \delta\bm{B}_{1}\cdot\nabla j_{z\rm eq}    + \nu_\perp\nabla^2\delta F_{1},\label{F1linear}\\
&\rho\partial_t \Delta\Omega_\theta = \frac{M}{r} + \frac{R}{r} + \nu_\perp\frac{1}{r^3}\frac{\partial}{\partial r}\left(r^3\frac{\partial}{\partial r}\Delta\Omega_\theta\right)\label{flow_p_eq},
\end{align}
where $B_\theta=d\psi_{\rm eq}/dr$, $\bm{B}_{\rm eq} = B_z\vec{e}_z + \vec{e}_z\times\nabla\psi_{\rm eq} = B_z\vec{e}_z + B_\theta\vec{e}_\theta$, $F_0=F_{\rm eq} + \delta F_0$, $\bm{v}_0 = \bm{v}_{eq} + \delta \bm{v}_0 = (v_{eq} + \delta v_{\theta 0})\vec{e}_\theta=r\Omega_\theta\vec{e}_\theta$, and $\delta v_{\theta 0} = r\Delta\Omega_\theta$ (see also Appendix A for detail). The Maxwell and Reynolds stresses, $M$ and $R$, satisfy
\begin{align*}
&M = -\frac{m}{r}\Imag\left\{\delta\psi_1^*\delta j_{z1} - \delta\psi_1\delta j_{z1}^*\right\},\\
&R = \rho\frac{m}{r}\Imag\left\{\delta\phi_1^*\delta F_{1} - \delta\phi_1\delta F_{1}^*\right\}.
\end{align*}
\section{Plasma response solutions with time dependent flow in RI and VR regimes}
In this section, we extend the previous plasma response solutions in slab configuration~\cite{huang19} to cylindrical geometry in presence of time-dependent poloidal flow for both RI and VR regimes. This part of the work is also an extension to the previous work on linear plasma response solution in Ref.~\cite{fitz91a}, where the steady state flow instead of time-dependent flow is considered. The latter extension is more consistent with the quasi-linear plasma flow response model developed later in Sec.~\ref{sec:flow_model}, where the plasma flow evolves in response to RMP and is indeed time-dependent.
\subsection{Plasma response equations in inner and outer regions}
Neglecting the flow shear terms, the linearized governing equations for $\delta\psi_1$ and $\delta\phi_1$, i.e. Eqs.~\eqref{psi1linear} and ~\eqref{F1linear}, can be reduced to
\begin{eqnarray}
&&\left(\frac{\partial}{\partial t} + im\Omega_s\right) \delta\psi_1 + \delta\bm{v}_1 \cdot \nabla \psi_{\rm eq} - B_z\partial_z \delta\phi_1 = \eta \delta j_{z1} \label{ln_psi},\\
&&\rho \left(\frac{\partial}{\partial t} + im\Omega_s\right) \delta F_1 = \bm{B}_{\rm eq} \cdot \nabla \delta j_{z 1} + \delta\bm{B}_1 \cdot \nabla j_{z\rm  eq}+\nu_\perp\nabla^2\delta F_1\label{ln_phi},
\end{eqnarray}
where $\Omega_s=\Omega_\theta(r_s,t)$, $r_s$ represents the $m/n$ rational surface. Note that $\Omega_s$ in Eqs.~\eqref{ln_psi} and \eqref{ln_phi} can be time-dependent.


In the outer region, Eq. \eqref{ln_phi} becomes
\begin{equation}
\frac{B_z}{R_0}\left(\frac{1}{q}-\frac{1}{q_s}\right)\delta j_{z1} = \frac{1}{r}\frac{dj_{ z\rm eq}}{dr}\delta\psi_1,
\end{equation}
where $q = rB_z/(R_0B_\theta)$ is the safety factor and $q_s = m/n$. Following Refs. \onlinecite{fitz00, huang15a}, we define $\delta\psi_1 (r, t) =\delta\psi_{1s}(r)\psi_s(t)+\delta\psi_{1c}(r)\psi_c(t)$, where $\psi_c=\delta\psi_1(a, t)$ and $\psi_s(t)=\delta\psi_1(r_s, t)$ represent the external RMP field and the corresponding plasma response in magnetic field on the resonant flux surface, respectively. Besides, $\delta\psi_{1s}$ and $\delta\psi_{1c}$ satisfy that
\begin{align}
 \delta\psi_{1s}(r_s, t)=1, \delta\psi_{1s}(a, t)=0, \delta\psi_{1c}(r_s, t)=0, \delta\psi_{1s}(a, t)=1.
\end{align}
Then, the index $\Delta' = [d\ln{\delta\psi_1}/dx]_{r_s}$ can be rewritten as $\psi_s\Delta' = \Delta_0'\psi_s + \Delta'_c \psi_c = [d\delta\psi_{1s}/dx]_{r_s}\psi_s + [d\delta\psi_{1c}/dx]_{r_s}\psi_c$, where $[f]_{r_s}\equiv f(r_s+)-f(r_s-)$ is the jump across the resonant flux surface at $r=r_s$.

To proceed, we define $ \hat\psi_1 \equiv \delta\psi_1 e^{i\varphi_{\rm temp}(t)}$ and $\hat\phi_1 \equiv \delta\phi_1 e^{i\varphi_{\rm temp}(t)}$, where $\varphi_{\rm temp} \equiv \int_0^t m\Omega_s(t') dt'$~\cite{huang19}. In the inner region, assuming $\partial_x \gg (m/r, n/R_0$), one can simplify Eqs. \eqref{ln_psi} and \eqref{ln_phi} as
\begin{eqnarray}
&&\frac{\partial \hat\psi_1}{\partial t} + i\frac{B_z}{R_0}C_0x\hat\phi_1 = \eta \hat j_{z 1} \label{li_in_psi},\\
&&\rho\frac{\partial\hat F_1}{\partial t} = - i\frac{B_z}{R_0}C_0x \hat j_{z 1}-i\frac{m}{r}\frac{dj_{z\rm eq}}{dr}\hat\psi_1+\nu_\perp\frac{\partial^2\hat F_1}{\partial x^2} \label{li_in_phi},
\end{eqnarray}
where $x = r- r_s$, $C_0 =mq'_s/q_s^2$. We further Laplace transform Eqs. \eqref{li_in_psi} and \eqref{li_in_phi}~\cite{hahm85} and neglect the second term on the right hand side of Eq. \eqref{li_in_phi}~\cite{furth63a}. Then, one arrives at
\begin{eqnarray}
&&\frac{\partial^2}{\partial \chi^2}\Psi = \delta \Omega\Psi_s(\frac{\Psi}{\Psi_s} + \chi \xi)\label{lnps_in_lp},\\
&&\frac{\delta_{VR}^6}{\delta^6}\frac{\partial^4}{\partial \chi^4}\xi - \frac{\delta_{RI}^4}{\delta^4}\frac{\partial^2}{\partial \chi^2}\xi + \chi^2 \xi + \chi \frac{\Psi}{\Psi_s} = 0\label{lnph_in_lp},
\end{eqnarray}
where $\chi = x/\delta_{\rm layer}=x / (\delta r_s)$, $\delta_{\rm layer}\equiv\delta r_s$ is the resistive tearing layer width, and
\begin{eqnarray*}
&\tilde\psi = \mathcal{L}[\hat\psi_1]  = \displaystyle \int_0^\infty \hat\psi_1 e^{- s t} d t,\\
&\tilde \phi = \mathcal{L}[\hat\phi_1] = \displaystyle \int_0^\infty \hat\phi_1 e^{- s t} d t,\\
&\tau_R=\frac{\mu_0r_s^2}{\eta}, \tau_H=\frac{R_0}{B_z}\frac{\sqrt{\mu_0\rho}}{r_sC_0}, \tau_V=\frac{r_s^2\rho}{\nu_\perp},\\
&\delta_{RI}^4=\frac{s \tau_H^2}{\tau_R}, \delta_{VR}^6=\frac{\tau_H^2}{\tau_R\tau_V}, \nu = \frac{is}{C_0^2\epsilon_s \delta}, \Omega = \delta \tau_R s, \epsilon_s=\frac{r_s}{R_0},\\
&\Psi = \frac{C_0}{B_z} \tilde \psi, U = - \tilde \phi / \nu, \Psi_s = \Psi(r_s), \Psi_c = \Psi(a), \xi=U/\Psi_s.
\end{eqnarray*}

Equations similar to Eqs.~\eqref{lnps_in_lp} and~\eqref{lnph_in_lp} are first proposed in Ref.~\cite{fitz94a}, where the formulas of solutions are given for the steady states. Here, we extend previous results in Refs.~\cite{fitz94a, huang19} to the solutions of linear plasma response in cylindrical geometry with time-dependent rigid flow in both RI and VR regimes.

\subsection{Plasma response solution in the RI regime}

In the RI regime, $\delta=\delta_{RI} \gg \delta_{VR}$, i.e. $s \gg \tau_R^\frac13/(\tau_H^\frac23\tau_V^\frac23)$, which is equivalent to $t \ll (\tau_H^\frac23\tau_V^\frac23)/\tau_R^\frac13$, the viscosity term in Eq.~\eqref{lnph_in_lp} can be neglected~\cite{fitz94a}. In addition, the constant-$\psi$ assumption in the inner region is valid when $\delta\Omega \ll 1$, i.e. $s \ll 1/(\tau_R^\frac13\tau_H^\frac23)$, so that $\Psi/\Psi_s\approx  1$ but $\partial_\chi^2\Psi$ should be kept in Eqs.~\eqref{lnps_in_lp} and \eqref{lnph_in_lp}. Thus in the constant-$\psi$ RI regime, Eq.~\eqref{lnph_in_lp} reduces to
\begin{eqnarray}
\frac{\partial^2}{\partial \chi^2}\xi_{RI} - \chi^2 \xi_{RI} = \chi\label{lnph_in_lpRI},
\end{eqnarray}
where $\xi=\xi_{RI}$. Eq.~\eqref{lnps_in_lp} becomes
\begin{align}
\frac{\partial^2}{\partial \chi^2}\Psi = \delta_{RI}\Omega\Psi_s(1 + \chi \xi_{RI}),
\end{align}
in the inner region. After the asymptotic matching, one arrives at
\begin{align}
\alpha_1\frac{ \Omega}{r_s} \Psi_s = \Delta_0'\Psi_s + \Delta_c'\Psi_c\label{HKFLOWLaplaceRI},
\end{align}
where $\alpha_1=\int_{-\infty}^{\infty}(1+\chi\xi_{RI})d\chi \approx 2.12$~\cite{huang19, militello04a}.

Using the inverse Laplace transform, Eq. \eqref{HKFLOWLaplaceRI} can be transformed to
\begin{align}
&\psi_s(t)= -\frac{\Delta_c'}{\Delta_0'}e^{-i\varphi_{\rm temp}(t)}\int_0^tG_1(t-t')\psi_c(t')e^{i\varphi_{\rm temp}(t')}dt'\label{newHK},
\end{align}
where
\begin{align*}
&G_1(t) = \frac{1}{\tau_{RI}}\left\{-\frac45[P_Ae^{P_A \tau}  + P_Be^{P_B \tau}] - \frac{\lambda_1}{\sqrt{2}\pi} \displaystyle \int_0^\infty e^{- u \tau} \frac{u^{\frac54}}{(1 - \sqrt{2} \lambda_1 u^{\frac54} + \lambda_1^2 u^{\frac52})} d u\right\},
\end{align*}
$\tau = t/\tau_{RI}$, $\tau_{RI}=\tau_R^{\frac35} \tau_H^{\frac25}$, $\lambda_1 = -\alpha_1/(r_s\Delta_0')$, and $P_{A,B}=\lambda_1^{-\frac45}{\rm exp}(\pm 4\pi i/5)$. Other than the geometry-dependent factors such as $\lambda_1$ and $\Delta_c'/\Delta_0'$, the expression of $\psi_s$ in Eq.~\eqref{newHK} is nearly the same as in the slab geometry~\cite{huang19}.

\subsection{Plasma response solution in the VR regime}

In the VR regime~\cite{fitz94a}, $\delta=\delta_{VR} \gg \delta_{RI}$, i.e. $s \ll \tau_R^\frac13/(\tau_H^\frac23\tau_V^\frac23)$, which is equivalent to $t \gg (\tau_H^\frac23\tau_V^\frac23)/\tau_R^\frac13$, the second term on the left hand side of Eq.~\eqref{lnph_in_lp} can be ignored. Additionally, the constant-$\psi$ assumption is equivalent to $\delta\Omega \ll 1$, i.e. $s \ll \tau_V^\frac13/(\tau_R^\frac23\tau_H^\frac23)$. Then Eq.~\eqref{lnph_in_lp} can be simplified as
\begin{eqnarray}
\frac{\partial^4}{\partial \chi^4}\xi_{VR} + \chi^2 \xi_{VR} = -\chi\label{lnph_in_lpVR},
\end{eqnarray}
where $\xi=\xi_{VR}$. Eq.~\eqref{lnps_in_lp} becomes
\begin{align}
\frac{\partial^2}{\partial \chi^2}\Psi = \delta_{VR}\Omega\Psi_s(1 + \chi \xi_{VR}),
\end{align}
in the inner region. Asymptotic matching leads to the following relation
\begin{align}
\alpha_2\frac{ \Omega}{r_s} \Psi_s = \Delta_0'\Psi_s + \Delta_c'\Psi_c\label{HKFLOWLaplaceVR},
\end{align}
where $\alpha_2=\int_{-\infty}^{\infty}(1+\chi\xi_{VR})d\chi \approx 2.103$~\cite{procelli87, militello11a}. Similar to the RI regime, Eq. \eqref{HKFLOWLaplaceVR} can be inverse Laplace transformed to the following
\begin{align}
\alpha_2\frac{\delta_{VR}\tau_R}{r_s} [\frac{d}{dt}+im\Omega_s]\psi_s = \Delta_0'\psi_s + \Delta_c'\psi_c\label{HKFLOWVR},
\end{align}
which yields the following
\begin{align}
&\psi_s = -\frac{\Delta_c'}{\Delta_0'}e^{-i\varphi_{\rm temp}(t)}\int_0^tG_2(t-t')\psi_c(t')e^{i\varphi_{\rm temp}(t')}dt'\label{HKFLOWVRsolution},
\end{align}
where
$G_2(t)=\frac{P_C}{\tau_{VR}}e^{-P_C\tau}$, $\tau=t/\tau_{VR}$, $\tau_{VR}=\delta_{VR}\tau_R$, $\lambda_2=-\alpha_2/(r_s\Delta_0')$, and $P_C=\lambda_2^{-1}$.

Note that Eq.~\eqref{HKFLOWVR} has also been heuristically derived by Fitzpatrick~\cite{fitz14} and further investigated by Beidler et al.~\cite{beidler17a, beidler18a}. As previously claimed in the appendix of Ref.~\cite{fitz14}, such an equation is derived from the following relations in Ref.~\cite{fitz98a}
\begin{align}
&\omega=m\Omega_\theta(r_s)-n\Omega_\phi(r_s)\label{fitz_1},\\
&P=\frac{\tau_R}{\tau_V}, Q=\tau_H^{\frac23}\tau_R^{\frac13}\omega\label{fitz_2},\\
&\delta_{VR}^F=\frac{\tau_H^{\frac13}}{\tau_R^{\frac16}\tau_V^{\frac16}}r_s, \hat\Delta=\Delta\frac{\delta_{VR}^F}{r_s}\label{fitz_3},\\
&\hat\Delta=-2.104e^{-i\pi/2}P^{\frac13}Q\label{fitz_4},
\end{align}
where the index $\Delta=\Delta(\omega)$ is also conventionally named as $\Delta'$ in our work. The rest of definitions are conventional and can be found following Eq. $(17)$
of Ref.~\cite{fitz98a}. It should be noted that the relations in above Eqs.~\eqref{fitz_1}-\eqref{fitz_4} are meant for steady state. In particular, Eq.~\eqref{fitz_4} is the steady state solution of Eq.~\eqref{HKFLOWLaplaceVR}. Thus, Eq. $(8)$ in Ref.~\cite{fitz14} does not directly derive from Eqs.~\eqref{fitz_1}-\eqref{fitz_4}.
In this work, we transfer the effect of time-dependent flow into the phases of $\hat\psi_1$ and $\hat\phi_1$, thus the resulting linearized reduced MHD equations \eqref{li_in_psi} and \eqref{li_in_phi} can be solved using Laplace transform as before~\cite{hahm85, huang19}. Our linear response solutions with time-dependent flow can be straightforwardly extended to various parameter regimes.
\section{Quasi-linear plasma flow response to external magnetic perturbation}
\label{sec:flow_model}
In previous section, we have developed a systematic derivation of the plasma response solutions with time-dependent flow in the RI an VR regimes. To close the plasma flow response model, the quasi-linear equations for the flow evolution are further derived in this section.
\subsection{The quasi-linear angular momentum equation in the Bessel function spectral space}
We expand the poloidal angular velocity as $\Delta\Omega_\theta = \sum_{k=0}^{\infty} a_k J_1(\mu_k \hat r)/\hat r$, where $\hat r=r/a$, $J_1$ is the first order Bessel function, and $\mu_k$ are the $k$-th zero points of $J_1$. Then, Eq. \eqref{flow_p_eq} can be transformed to
\begin{align}
&\rho\partial_t a_k = C_kM_k + C_kR_k - \frac{\nu_\perp}{a^2}\mu_k^2a_k \label{omegageneral},\\
&M_k=\frac{2m}{J_1(\mu_k\hat r_s)}\Imag\int_0^1 J_1(\mu_k\hat r)\left\{\delta\psi_1\delta j_{z1}^*\right\}d\hat r \label{Mgeneral},\\
&R_k=\frac{2m\rho}{J_1(\mu_k\hat r_s)}\Imag\int_0^1 J_1(\mu_k\hat r) \left\{\delta\phi_1^*\delta F_1\right\}d\hat r \label{Rgeneral},
\end{align}
where $C_k=\frac{J_1(\mu_k\hat r_s)}{a^2N_k}$ and $N_k = \frac{1}{2}J_2^2(\mu_k)$.

Since $\delta j_{z1}$, $\delta\phi_1$, and $\delta F_1$ are localized around the rational surface, the Maxwell and Reynolds stresses should be nonzero only in the inner region. Using Taylor expansion at the rational surface, $M_k$ and $R_k$ can be approximated as
\begin{align}
M_k &= \frac{2m}{J_1(\mu_k\hat r_s)a}\Imag\int_{r_{s-}}^{r_{s+}} J_1(\mu_k\hat r)\delta\psi_1\delta j_{z1}^*dr \nonumber\\
    &=\frac{2m}{J_1(\mu_k\hat r_s)a}\Imag\int_{r_{s-}}^{r_{s+}} [J_1(\mu_k\hat r_s)+\mu_kJ'_1(\mu_k\hat r_s)x]\delta\psi_1\delta j_{z1}^*dr \nonumber\\
    &=F_m + D_k N_m, \label{Maxwell_MG}
\end{align}
and
\begin{align}
R_k &= \frac{2m\rho}{J_1(\mu_k\hat r_s)a}\Imag\int_{r_{s-}}^{r_{s+}} J_1(\mu_k\hat r)\delta\phi_1^*\delta F_1dr \nonumber\\
    &=\frac{2m\rho}{J_1(\mu_k\hat r_s)a}\Imag\int_{r_{s-}}^{r_{s+}} [J_1(\mu_k\hat r_s)+\mu_kJ'_1(\mu_k\hat r_s)x]\delta\phi_1^*\delta F_1dr \nonumber\\
    &=F_r + D_k N_r, \label{Reynolds_MG}
\end{align}
where $D_{k}=\frac{\mu_kJ'_1(\mu_k\hat r_s)}{J_1(\mu_k\hat r_s)}$, and
\begin{align}
&F_m = \frac{2m}{a}\Imag\int_{r_{s-}}^{r_{s+}} \delta\psi_1\delta j_{z1}^*dr\label{Maxwell_Force},\\
&F_r = \frac{2m\rho}{a}\Imag\int_{r_{s-}}^{r_{s+}} \delta\phi_1^*\delta F_1dr\label{Reynolds_Force},\\
&N_m = \frac{2m}{a}\Imag\int_{r_{s-}}^{r_{s+}} x\delta\psi_1 \delta j_{z1}^*dr \label{MaxwellG},\\
&N_r = \frac{2m\rho}{a}\Imag\int_{r_{s-}}^{r_{s+}} x\delta\phi_1^*\delta F_1dr\label{MaxwellG}.
\end{align}
Similar to Ref.~\cite{cole15a}, we define the quasi-linear forces and moments as the zeroth and first moments of the relevant stresses, where $F_m$ and $F_r$ are the Maxwell and Reynolds forces, and $N_m$ and $N_r$ are Maxwell and Reynolds moments, respectively. They are the lowest and the next order terms in Taylor expansion series of the relevant stresses in the Bessel spectral space.

 Combining Eqs.~\eqref{omegageneral},~\eqref{Maxwell_MG}, and \eqref{Reynolds_MG}, one obtains the following equation
\begin{align}
&\rho\partial_t a_k = C_k(F_m + F_r + D_kN_m + D_kN_r) - \frac{\nu_\perp}{a^2}\mu_k^2a_k\label{bessel}.
\end{align}
The above equation reduces to the previous poloidal torque balance equation in Bessel spectral space if one neglects $F_r$, $N_r$, and $N_m$~\cite{huang15a, fitz2019}. Different from the conventional model in Ref.~\cite{fitz93a}, the torque balance equation in real space is not needed, nor is any assumption on the radial profile of Maxwell torque. In contrast, we construct the quasi-linear plasma flow model from the plasma response solution and the poloidal angular momentum equation in the Bessel spectral space, which can also be extended to include the toroidal flow. On the other hand, Cole et al.~\cite{cole15a} argue that the forces tend to cause the tearing mode locking whereas the moments determine the evolution of the flow shear. From Eq.~\eqref{bessel}, one finds that the plasma flow as well as its shear can be modified by both the forces and the moments.

We further neglect the current gradient and flow shear terms, and adopt the constant-$\psi$ assumption. Since $\delta j_{z1}$ is an even function of $x$, and $\delta\phi_1$ and $\delta F_1$ are odd functions of $x$~\cite{militello11a, cole15a}, one can simplify Eqs.~\eqref{Maxwell_Force}-\eqref{MaxwellG} as
\begin{align}
&F_m = \frac{2m}{\mu_0a}|\psi_s|^2\Imag{\Delta'}=\frac{2m}{\mu_0a}|\psi_s|^2\Delta_c'\Imag{\left\{\frac{\psi_c}{\psi_s}\right\}}\label{Maxwell_Force2},\\
&F_r = \frac{2m\rho}{a}\Imag\int_{r_{s-}}^{r_{s+}} \delta\phi_1^*\delta F_1dr\label{Reynolds_Force2},\\
&N_m = 0,\\
&N_r = 0.
\end{align}
Note that the quasi-linear moments in this work are exactly zero. When effects such as flow shear in the inner region are considered, the plasma flow evolution could be significantly influenced by the Maxwell and Reynolds moments~\cite{cole15a}.

\subsection{Quasi-linear forces in the RI and VR regimes}
In the steady state RI regime with constant-$\psi$ assumption, $\psi_s$ satisfies the following equation
\begin{align}
\alpha_1\frac{im\Omega_{s}\delta_{RI}\tau_R}{r_s}\psi_s = \Delta_0'\psi_s + \Delta_c'\psi_c\label{psiRI}.
\end{align}
Combing Eq.~\eqref{psiRI} and the relationship between $\delta\phi_1$ and $\psi_s$~\cite{huang19}, the steady state Maxwell and Reynolds forces can be written as
\begin{align}
&F_m = \frac{2m}{\mu_0a}|\psi_s|^2\Delta_c'\Imag{\left\{\frac{\psi_c}{\psi_s}\right\}}=\frac{2m}{\mu_0a}|\psi_s|^2\frac{\alpha_1}{r_s}|m\Omega_{ s}\tau_{RI}|^{\frac54}\sgn(m\Omega_s)\sin\frac58\pi\label{Maxwell_ForceRI},\\
&F_r = \frac{2m\rho}{a}\Imag\int_{r_{s-}}^{r_{s+}} \delta\phi_1^*\delta F_1dr=-\frac{2m}{\mu_0a}|\psi_s|^2\frac{\alpha_3}{r_s}|m\Omega_{ s}\tau_{RI}|^{\frac54}\sgn(m\Omega_s)\sin\frac18\pi\label{Reynolds_ForceRI},
\end{align}
where $\alpha_1 = \int_{-\infty}^{\infty}[1+\chi\xi_{RI}]d\chi \approx 2.12$ and $\alpha_3=\int_{-\infty}^{\infty}\xi_{RI}\partial_\chi^2\xi_{RI}d\chi \approx 0.54$. Obviously, the Reynolds force $F_r$ is opposite sign to the Maxwell force $F_m$ and $F_r < F_m$. Note that the ratio of $F_r/F_m$ is a constant independent of equilibrium, which is similar with the case in slab geometry~\cite{huang19}.

On the other hand, within constant-$\psi$ assumption, $\psi_s$ in the steady state VR regime satisfies
\begin{align}
\alpha_2\frac{im\Omega_s\delta_{VR}\tau_R}{r_s}\psi_s = \Delta_0'\psi_s + \Delta_c'\psi_c\label{psiVR}.
\end{align}
Similar to the case in the steady state RI regime, the Maxwell and Reynolds forces can be written as
\begin{align}
&F_m = \frac{2m}{\mu_0a}|\psi_s|^2\Delta_c'\Imag{\left\{\frac{\psi_c}{\psi_s}\right\}}=\frac{2m}{\mu_0a}|\psi_s|^2\frac{\alpha_2}{r_s}|m\Omega_s\tau_{VR}|\sgn(m\Omega_s)\sin\frac12\pi\label{Maxwell_ForceVR},\\
&F_r = \frac{2m\rho}{a}\Imag\int_{r_{s-}}^{r_{s+}} \delta\phi_1^*\delta F_1dr=0\label{Reynolds_ForceVR}.
\end{align}
Note that the Reynolds force in the steady state VR regime is exactly zero. In fact, the above expressions for the Reynolds force in the steady state RI and VR regimes differ from previous results~\cite{cole15a}. This is because our Reynolds force is defined from $\delta\phi_1$ and $\delta F_1$ in the inner region, whereas the $F_r$ in the previous work is calculated using the approximated expressions of perturbed stream function and vorticity in the outer region. On the other hand, the above expressions of the Maxwell force in both the steady state RI and VR regimes recover the previous results except the geometry factors~\cite{cole15a}.
\subsection{Analytical plasma flow model in presence of RMP}
When the island width $W$ is still much smaller than the resistive tearing layer width $\delta_{\rm layer}$, quasi-linear magnetic terms may be neglected. Based on the Laplace transform and Bessel expansion, we propose a rigorous derivation for the plasma flow model in response to RMP in cylindrical geometry. In this model, the island evolution is determined by the linear plasma response solutions in the constant-$\psi$ RI and VR regimes, i.e. Eqs.~\eqref{newHK} and \eqref{HKFLOWVRsolution}, which are used to obtain the equation for plasma flow in the Bessel spectral space. We further demonstrate that quasi-linear moments are exactly zero and Reynolds force can always be neglected. We summarize the model for plasma flow below
\begin{align}
&\Omega_s=\sum_{k=0}^{\infty}a_k\frac{J_1(\mu_k\hat r_s)}{\hat r_s}+\Omega_{\rm eq}(r_s)\label{mod1},\\
&\rho\partial_t a_k = C_kF_m - \frac{\nu_\perp}{a^2}\mu_k^2a_k,\\
&F_m = \frac{2m}{\mu_0a}|\psi_s|^2\Delta_c'\Imag{\left\{\frac{\psi_c}{\psi_s}\right\}}\label{mod3},
\end{align}
where $C_k=\frac{J_1(\mu_k\hat r_s)}{a^2N_k}$, and $N_k=\frac12J_2(\mu_k^2)$.

Different from previous work, the above equations~\eqref{mod1}-\eqref{mod3} are naturally derived from the two-field reduced MHD model. Note that the plasma response solutions in the RI and VR regimes, i.e. Eqs.~\eqref{newHK} and ~\eqref{HKFLOWVRsolution}, are appropriate only in the small island regime. When $W \gg \delta_{\rm layer}$, quasi-linear magnetic terms should be included.
\section{Summary and discussion}
In summary, we have developed an analytical model for the quasi-linear plasma flow response to RMP in the RI and VR regimes within the framework of two-field reduced MHD equations. Neglecting the quasi-linear magnetic effects, previous linear solutions on plasma response in magnetic field~\cite{fitz91a,huang19} are extended to cylindrical geometry in presence of time-dependent rigid poloidal flow for both RI and VR regimes. The extension to the linear plasma response solutions in presence of steady state flow~\cite{fitz91a} is to allow time-dependent flow, which is more consistent with the quasi-linear plasma flow response model developed in this work, where the plasma flow evolves in response to RMP and is indeed time-dependent. The corresponding plasma flow response equation including the quasi-linear forces and moments is derived in the Bessel spectral space, without invoking any assumption on the Maxwell torque or its radial profile. Different from previous works, our analytical model is built purely from the two-field reduced MHD equations, which allow us to accurately calculate and clarify the physical origins of the quasi-linear forces and moments self-consistently.

Due to the limitation of the two-field reduced MHD model, many physics elements for the RMP induced plasma response have not been included. For example, two-fluid, neo-classical, finite-orbit-width, and finite-Larmor-radius effects are known to have strong influence over tearing modes as well as plasma response to RMPs near resonant surfaces\cite{fitz98a, wael12, fitz2019, zhang2020}. Furthermore, our derivation is appropriate only in the small island regime ($W \ll \delta_{\rm layer}$). We plan to address these important issues in future studies.

\begin{acknowledgments}
This work was supported by the Fundamental Research Funds for the Central Universities at Huazhong University of Science and Technology Grant No. 2019kfyXJJS193, the National Natural Science Foundation of China Grant No. 11775221, 11763006, and 51821005, the Young Elite Scientists Sponsorship Program by CAST Grant No. 2017QNRC001, and U.S. Department of Energy Grant Nos. DE-FG02-86ER53218 and DE-SC0018001.
\end{acknowledgments}

\section*{data availability}
Data sharing is not applicable to this article as no new data were created or analyzed in this study.

\newpage
\begin{appendix}
\section{Derivation of equation for $\Delta\Omega_\theta$}

We write any quantity as $f = f_{\rm eq} + \delta f$, where $f_{\rm eq} = f_{\rm eq}(r)$ and $\delta f = \delta f(r, \theta, \phi, t)$ are the equilibrium and perturbation parts of $f$. We expand the perturbed quantity as $\delta f = \sum_{l=-\infty}^\infty  \delta f_le^{il(m\theta-n\phi)}$. Using the single helicity assumption, the quasi-linear equation for the $0/0$ component of $\delta F$, i.e. $\delta F_0$, can be obtained from Eq.~\eqref{rmhd2}
\begin{align}
\rho\frac{\partial}{\partial t} \delta F_0 + \rho (\delta\vec{v}_1^*\cdot\nabla \delta F_1+\delta\vec{v}_1\cdot\nabla \delta F_1^*)=\delta\vec{B_1}\cdot\nabla \delta j_{z1}^* + \delta\vec{B_1}^*\cdot\nabla \delta j_{z1} + \nu_\perp \nabla_\perp^2 \delta F_0\label{A0},
\end{align}
where
\begin{align}
&\delta v_{\theta 0} = \frac{\partial}{\partial r}\delta\phi_0 = r\Delta\Omega_\theta,\\
&\delta F_0 = \nabla^2_\perp\delta\phi_0 = \frac{1}{r}\frac{\partial}{\partial r}(r\frac{\partial}{\partial r}\delta\phi_0) = \frac{1}{r}\frac{\partial}{\partial r}(r^2\Delta\Omega_\theta)\label{Aa},\\
&\nabla_\perp^2\delta F_0 = \frac{1}{r}\frac{\partial}{\partial r}\left\{r\frac{\partial}{\partial r}\delta F_0\right\}= \frac{1}{r}\frac{\partial}{\partial r}\left\{r\frac{\partial}{\partial r}[\frac{1}{r}\frac{\partial}{\partial r}(r^2\Delta\Omega_\theta)]\right\}=\frac{1}{r}\frac{\partial}{\partial r}\left\{\frac{1}{r}\frac{\partial}{\partial r}(r^3\frac{\partial}{\partial r}\Delta\Omega_\theta)\right\},\\
&\delta\vec{B_1}\cdot\nabla \delta j_{z1}^* + \delta\vec{B_1}^*\cdot\nabla \delta j_{z1} = \frac{im}{r}\frac{\partial}{\partial r}\left\{\delta\psi_1^*\delta j_{z1} - \delta\psi_1\delta j_{z1}^*\right\}=-\frac{m}{r}\frac{\partial}{\partial r}\Imag\left\{\delta\psi_1^*\delta j_{z1} - \delta\psi_1\delta j_{z1}^*\right\},\\
&\delta\vec{v_1}\cdot\nabla \delta F_{1}^* + \delta\vec{v_1}^*\cdot\nabla \delta F_{1} = \frac{im}{r}\frac{\partial}{\partial r}\left\{\delta\phi_1^*\delta F_{1} - \delta\phi_1\delta F_{1}^*\right\}=-\frac{m}{r}\frac{\partial}{\partial r}\Imag\left\{\delta\phi_1^*\delta F_{1} - \delta\phi_1\delta F_{1}^*\right\}\label{Ad}.
\end{align}
Here, higher harmonics are neglected in the quasi-linear approximation. Substituting Eqs.~\eqref{Aa}-\eqref{Ad} into Eq.~\eqref{A0} and performing integral $\int_0^r rdr$ on both sides of Eq.~\eqref{A0}, we obtain
\begin{align}
\rho\partial_t \Delta\Omega_\theta = \frac{M}{r} + \frac{R}{r} + \nu_\perp\frac{1}{r^3}\frac{\partial}{\partial r}(r^3\frac{\partial}{\partial r}\Delta\Omega_\theta),
\end{align}
where
\begin{align*}
&M = -\frac{m}{r}\Imag\left\{\delta\psi_1^*\delta j_{z1} - \delta\psi_1\delta j_{z1}^*\right\},\\
&R = \rho\frac{m}{r}\Imag\left\{\delta\phi_1^*\delta F_{1} - \delta\phi_1\delta F_{1}^*\right\}.
\end{align*}

\end{appendix}

\newpage

\bibliographystyle{apsrev}


\end{document}